# Microsecond Time-Resolved Cryo-Electron Microscopy

Ulrich J. Lorenz[*]

**Affiliation:** Ecole Polytechnique Fédérale de Lausanne (EPFL), Laboratory of Molecular Nanodynamics, CH-1015 Lausanne, Switzerland

* To whom correspondence should be addressed. Email: ulrich.lorenz@epfl.ch

**Abstract**

Microsecond time-resolved cryo-electron microscopy has emerged as a novel approach for directly observing proteins dynamics. By providing microsecond temporal and near-atomic spatial resolution, it has the potential to elucidate a wide range of dynamics that were previously inaccessible and therefore, to significantly advance our understanding of protein function. This review summarizes the properties of the laser melting and revitrification process that underlies the technique and describes different experimental implementations. Strategies for initiating and probing dynamics are discussed. Finally, the microsecond time-resolved observation of the capsid dynamics of CCMV, an icosahedral plant virus, is reviewed, which illustrates important features of the technique as well as its potential.

**Introduction**

Protein structure determination has made remarkable progress, particularly with the recent success of cryo-electron microscopy (cryo-EM).[1] Machine learning based approaches have even made it possible to predict protein structure from the amino acid sequence with reasonable confidence.[2,3] In contrast, our understanding of protein function is lagging behind, since it is not routinely possible to observe proteins as they perform their tasks.[4] Predicting their function from the amino acid sequence currently appears out of reach entirely.[5] Observing proteins at work is challenging because it requires not only near-atomic spatial resolution, but also a time resolution of at least microseconds, the timescale on which the large-amplitude domain motions occur that are typically associated with the activity of a protein.[6] Among the available time-resolved techniques that offer atomic resolution, ultrafast x-ray crystallography[7–9] requires the proteins to be packed in a crystal, which hinders many large-amplitude motions,[4] while NMR spectroscopy[4] and traditional time resolved cyro-EM[10] are usually several orders of magnitude too slow to observe many relevant processes. Over the last years, microsecond time-resolved cryo-EM has emerged as a novel technique that promises to fill this gap by enabling observations of protein dynamics, with both microsecond temporal as well as near-atomic spatial resolution.

**Concept**

Figure 1 illustrates the concept of microsecond time-resolved cryo-EM experiments. A cryo sample is flash melted with a laser pulse, and dynamics of the embedded proteins are initiate with a suitable stimulus as soon as the sample is liquid. While these dynamics unfold, the laser, which keeps the sample liquid, is switched off at a well-defined point in time. The sample cools rapidly and revitrifies, trapping the proteins in their transient configurations, which can subsequently be reconstructed with single particle cryo-EM techniques. As detailed in the following, this approach has several crucial features that make it suitable for the observation of protein dynamics.

**Instrumentation and sample geometry**

Melting and revitrification experiments were first performed *in situ*, that is inside a transmission electron microscope, as illustrated in Fig. 1a.[11,12] Microsecond laser pulses are created by chopping the output of a continuous wave laser (532 nm) with an acousto-optic modulator. The laser beam enters the microscope column from the left, is reflected by a small mirror above the upper pole piece of the objective lens, and strikes the sample at close to normal incidence. Figure 1b illustrates the sample geometry. Holey gold specimen supports on a gold mesh are well suited because of their high heat conductivity. The laser beam is focused onto the center of a grid square (typical spot size of tens of microns), where it locally melts the cryo sample by heating up the gold film. In a typical experiment, a circular area with a diameter of 5–10 holes is revitrified. As illustrated in Fig. 2b, the surrounding sample areas, which are too far from the center of the laser focus and do not reach the melting point, crystallize.[13] Conveniently, the appearance of a crystalline ring around the revitrified area provides on-the-fly feedback on the success of a revitrification experiment. Moreover, by adjusting the laser power to keep the diameter of the revitrified area constant, a repeatable temperature evolution across different grid squares can be ensured.

Alternatively, microsecond time-resolved cryo-EM experiments can also be carried out with a correlative light-electron microscopy approach.[14] As illustrated in Fig. 1c, revitrification is performed in an optical microscope, with the sample held in a liquid nitrogen cooled cryo stage. As before, the melting laser is focused onto the sample, striking it at normal incidence. After several grid squares have been revitrified, the sample is then transferred to an electron microscope for high-resolution imaging. This correlative approach will certainly be more accessible to most researchers since it is technically less involved than the *in situ* experiments. A drawback is that it is more difficult to obtain on-the-fly feedback on the success of an experiment, particularly since the crystalline ring surrounding the revitrified area is more challenging to visualize.

**Temperature evolution of the sample and time resolution**

The characteristic temperature evolution of the sample under laser irradiation determines the time resolution of the experiment. A simulation is shown in Fig. 3a.[15] For a bare specimen support (red curve), the temperature in the center of the laser focus rises rapidly after the laser is switched on, before it plateaus and remains stable for the remainder of the laser pulse. Once the laser is switched off, heat is efficiently dissipated towards the surroundings, which have remained at cryogenic temperature, and the sample cools rapidly (1 µs time constants for heating and cooling). The simulation agrees well with the experimental characterization of the temperature evolution shown in Fig. 3b, where, the diffraction intensity of the gold film (reflections highlighted in the diffraction pattern in the inset) is used as a temperature probe and is stroboscopically recorded with nanosecond electron pulses.[11] The simulations in Fig. 3a show that increasing the thickness of the vitreous ice layer adds more inertia to the system, with both the heating and cooling times increasing to a few microseconds (blue-green curves). This is consistent with time-resolved electron diffraction experiments that directly probe the temperature evolution of cryo sample.[16] Importantly, the speed with which the sample cools dictates how rapidly transient states can be trapped and therefore determines the time resolution. For a typical sample geometry and ice thicknesses suitable for cryo-EM, one can therefore achieve a time-resolution of 5 µs or better for typical gold specimen supports.[15]

Note that the simulations in Fig. 3a assume that the sample does not evaporate. In the vacuum of an electron microscope, this is not the case unless evaporation is prevented, for example by sandwiching the sample between two graphene layers.[15] If evaporation occurs, the temperature evolution remains qualitatively similar, even though higher laser powers are required to compensate for the evaporative cooling of the sample.[16] Evaporative cooling is useful by providing a negative feedback that limits the maximum temperature that the sample can reach, close to room temperature for typical sample geometries.[13] At the same time, the evaporation and eventual breakup of the liquid film limit the maximum observation time to tens of microseconds.

**Spatial resolution**

The melting and revitrification process leaves the particles intact and does not appear to impose a limit on the attainable spatial resolution. This is demonstrated in Fig. 3c, which compares reconstructions of apoferritin from a conventional and an *in-situ* revitrified cryo sample (details in Fig. 3d).[17] Within the resolution of the reconstructions of 1.6 Å, the two structures are indistinguishable. A similar result is obtained with the correlative approach.[14] Evidently, the melting and revitrification process does not damage the particles, nor does it induce any structural changes. This can be rationalized by considering that flash melting simply reverses the vitrification process, which is well-known to preserve protein structure. A notable difference is that during laser melting with a rectangular laser pulse, about a third of the sample transiently crystallizes.[18] If necessary, this can be avoided by adding an intense initial spike to the rectangular laser pulse and so achieve a heating rate in excess of $10^8$ K/s, which is sufficient to outrun crystallization.[19]

**Properties of revitrified cryo samples and overcoming preferred orientation**

While the melting and revitrification process leaves the structure of the embedded particles unchanged, it does somewhat alter the properties of the cryo sample. While only minor changes are observed in the beam-induced specimen motion,[20] the angular distribution of the particles is significantly altered.[17] As shown in Fig. 3e, a conventional apoferritin sample exhibits distinct maxima in the angular distribution, which result from the preferential adsorption of hydrophobic parts of the protein surface to the air water interface.[21,22] Unlike apoferritin, some asymmetric particles may show such severe preferred orientation that the number of available views becomes too limited to obtain a reconstruction — a common reason for cryo-EM projects to fail. Interestingly, revitrification produces a more even angular distribution (Fig. 3f), which is also observed for other proteins.[23] Evidently, small forces are exerted upon the particles during the melting and revitrification process, which reshuffles their angular distribution. This is in line with the observation that revitrified samples frequently also exhibit an uneven spatial distribution of the particles. Melting and revitrification may therefore potentially provide a simple tool for overcoming issues with preferred orientation.

**Initiating and observing dynamics**

A range of stimuli for initiating protein dynamics are readily available that are compatible with the melting and revitrification approach. For example, the melting laser itself can serve as a stimulus if its power is increased to induce a temperature jump.[13] It is also possible to use a second laser pulse to trigger a photoactive protein once the sample is liquid.[24,25] One of the most versatile approaches consists in using light to release photocaged reagents, such as caged ATP, ions, redox active compounds, or small peptides.[26–28]

Figure 4a–e illustrates the implementation of such a time-resolved experiment,[29] which uses a caged proton to trigger the dynamics of cowpea chlorotic mottle virus (CCMV)[30,31]. During its life cycle, this icosahedral plant virus switches between a contracted and an extended configuration, a motion that involves large-amplitude movements of the capsid proteins and that can be initiated by a pH jump.[32] As illustrated in Fig. 4a, a cryo sample of the extended state is prepared in the presence of a caged proton. With the sample still in its vitreous state, the caged proton is then released through UV irradiation, which drops the pH from 7.6 to 4.5. At such a low pH, the contracted configuration is most stable (Fig. 4e, as obtained from a sample prepared at pH 5.0). However, the particles cannot react to the pH jump as long as they are trapped in the matrix of vitreous ice, and contraction only begins once the sample is melted (Fig. 4b,c). When the laser is switched off after 30 µs and the sample revitrifies, the particles are trapped in partially contracted configurations (Fig. 4d).

A more detailed analysis reveals that the transient ensemble obtained after a 30 µs features significant conformational heterogeneity. Figure 4f shows a variability analysis (cryoSPARC[33]) of the mixed datasets of the extended, partially contracted, and fully contracted particles. By slicing the particle distribution along the first variability component and obtaining reconstructions for each slice, details of the contraction mechanism can be revealed. Figure 4g shows that the partially contracted ensemble (slices 8–12) features a large spread in particle diameters. This is because the contraction occurs in a dissipative medium, so that the particles move at different speeds. The reconstruction of Fig. 4c,d therefore corresponds to a motion-blurred average and consequently only yields a comparably low resolution of 8.0 Å. The contraction is accompanied by rotations of the pentamers and hexamers of the capsid as well as superimposed rotations of the capsid subunits (Fig. 4g). The observation of the capsid dynamics of CCMV demonstrates for the first time that microsecond time-resolved cryo-EM enables the study of protein dynamics that occur *in vivo*. It is difficult to imagine how any other approach could currently yield a similarly detailed picture of the mechanics of this nanoscale machine.

**Outlook**

The experiments reviewed here establish all the crucial features of the melting and revitrification approach to make it possible to directly observe a wide range of protein dynamics that have previously been inaccessible. Several technical challenges remain to be addressed. During laser melting, evaporation and eventual breakup of the thin liquid film currently limit the temporal observation window of the technique to several tens of microseconds. In order to bridge the gap to traditional, millisecond time-resolved cryo-EM experiments,[10] it is desirable to extend this window to several hundreds of microseconds. At the same time, it should also be possible to probe even faster dynamics on the nanosecond timescale. Shaped laser pulses that offer higher heating rates are a first step to improve the time resolution of the technique.[19] The large conformational heterogeneity of the transient ensemble observed for CCMV is likely going to be a general feature of microsecond time-resolved cryo-EM experiments. While this has made it possible to capture a large fraction of the contraction trajectory of CCMV from just a single time point, it simultaneously presents a challenge for the data analysis. Computational approaches for conformational sorting[34,35] will therefore likely play a crucial role in analyzing time-resolved experiments. It should be noted that as a single-particle technique, cryo-EM offers inherent advantages for dealing with such dynamic heterogeneity compared with x-ray crystallography or NMR spectroscopy. Finally, it will be crucial to make the technology easily accessible for the broader community. This makes it an urgent goal to refine and automate the correlative approach discussed here.[14] If watching proteins perform their function becomes indeed straightforward and routine, large numbers of such observations may eventually form the basis for training machine learning algorithms to predict function or even design artificial enzymes.[5]

**Declaration of competing interest**

The author declares that he has no known competing financial interests or personal relationships that could have appeared to influence the work reported in this paper.

**Data availability**

This is a review article.

**Acknowledgments**

This work was supported by the Swiss National Science Foundation Grants PP00P2_163681 and 200020_207842.

**Annotated references**

• of special interest

•• of outstanding interest

•• Harder OF, Barrass SV, Drabbels M, Lorenz UJ: **Fast viral dynamics revealed by microsecond time-resolved cryo-EM**. *Nat Commun* 2023, **14**:5649.

Observation of the microsecond motions of the capsid of CCMV. The pH jump experiment in this study can serve as a template for initiating a wide range of dynamics. The data analysis reveals several characteristics that are likely going to be general features of time-resolved experiments.

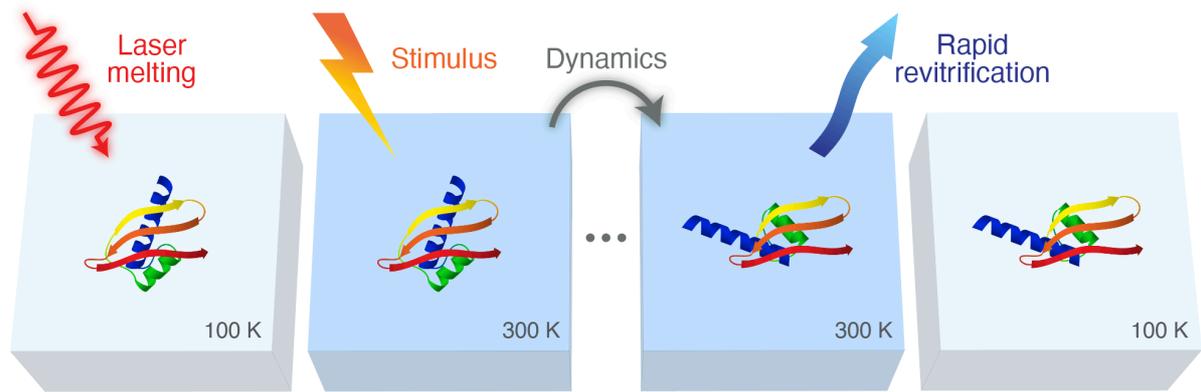

**Figure 1 | Experimental concept of microsecond time-resolved cryo-EM.** A cryo sample is melted with a microsecond laser pulse, and dynamics of the embedded particles are induced once the sample is liquid. As they unfold, the heating laser is switched off, and the sample rapidly cools and revitrifies, trapping proteins in their transient configurations. (Adapted from Refs. [13,15].)

**Sample geometry and instrumentation**

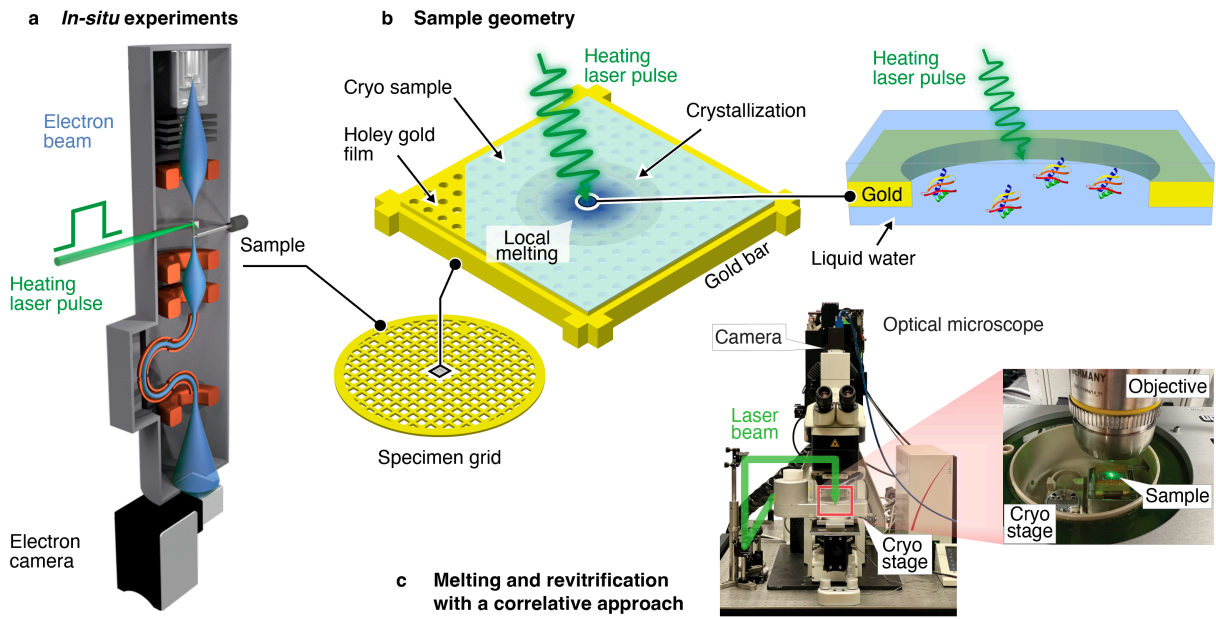

**Figure 2 | Sample geometry and instrumentation. a** Illustration of the modified transmission electron microscope for *in-situ* melting and revitrification experiments. (Adapted from Ref. [13].) **b** Illustration of the sample geometry. (Adapted from Ref. [16].) **c** Photograph of the optical microscope used for melting and revitrification experiments with a correlative light-electron microscopy approach. (Adapted from Ref. [14].)

**Temporal and spatial resolution as well as changes in preferred particle orientation**

**Temporal resolution**                                                    **Spatial resolution**

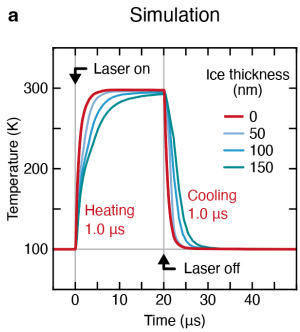 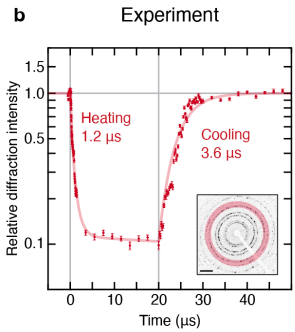 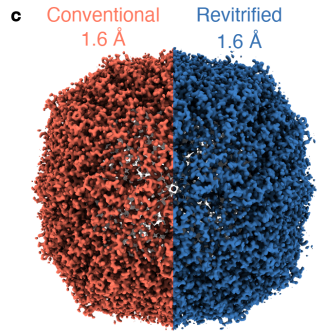 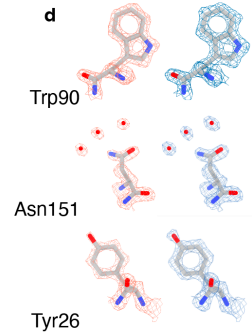

**Overcoming preferred orientation**

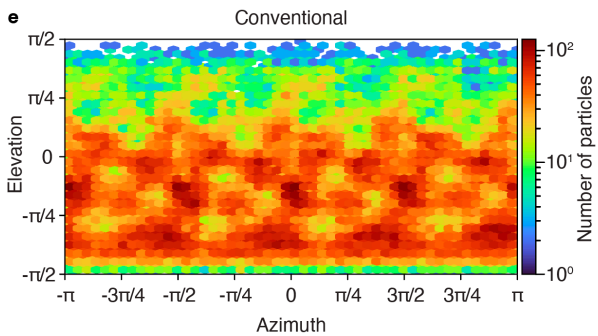 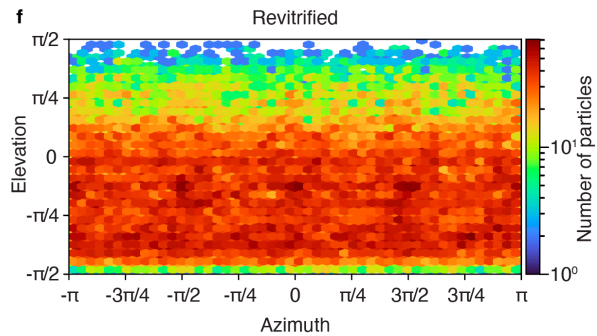

**Figure 3 | Temporal and spatial resolution of melting and revitrification experiments as well as changes in the preferred particle orientation. a** Simulation of the temperature evolution of a typical samples under laser irradiation. Note that evaporation cooling of the sample, which is not included here, provides an additional negative feedback on the temperature evolution. **b** Experimental characterization of the temperature evolution from a time-resolved diffraction experiment with nanosecond electron pulses. The diffraction intensity of the gold film is used as the temperature probe, with the select reflections highlighted in the diffraction pattern in the inset. Scale bar, 5 nm$^{-1}$. (Adapted from Ref. [15].) **c** Reconstructions of apoferritin from conventional and revitrified cryo samples are indistinguishable. **d** Details of the reconstructions in **c**, with a model of apoferritin[36] placed into the density through rigid-body fitting. **e,f** Revitrification reduces preferred particle orientation, as evident from the angular distribution of the particles in the reconstructions in **c**. (Adapted from Ref. [17].)

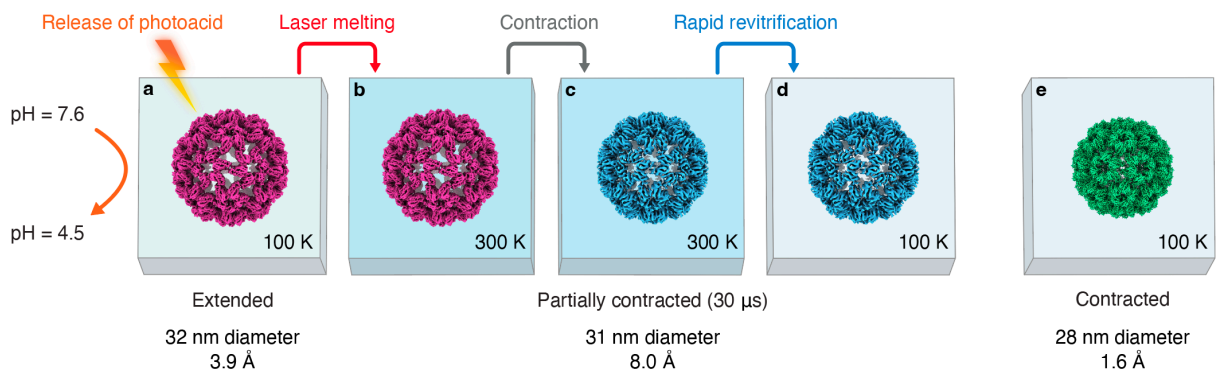

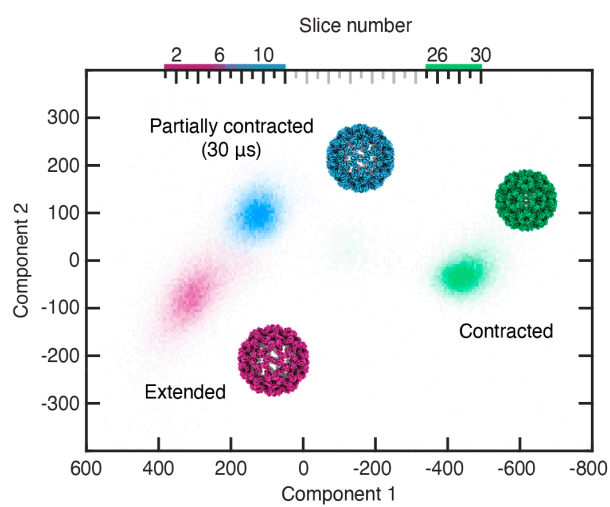
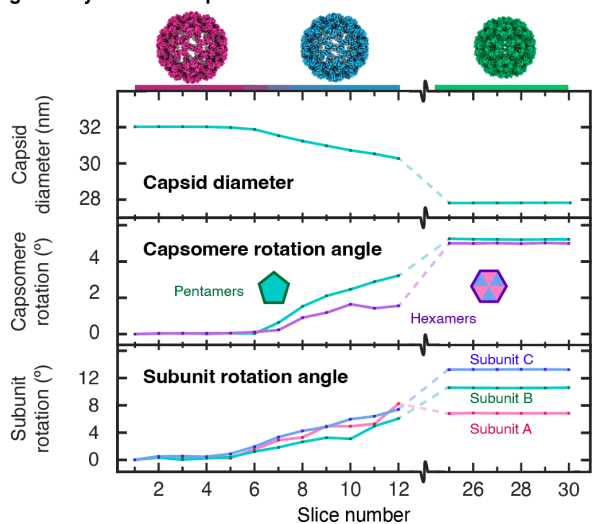

**Figure 4 | Observing protein dynamics on the microsecond timescale: Capsid motions of CCMV. a–e** Experimental concept. A cryo sample of the virus is prepared at pH 7.6, at which its capsid assumes an extended configuration (**a**). By releasing a photoacid, the pH is then jumped down to 4.5. At such a low pH, the contracted state of the capsid is most stable (**e**). However, contraction only begins once the sample is laser melted (**b,c**). Upon revitrification, partially contracted configurations are trapped (**d**). **f** Variability analysis of the extended, partially contracted, and fully contracted ensembles. **g** Analysis of the contraction mechanism of CCMV. The particle distribution in **f** is sliced along the first component, and reconstructions are obtained for each slice, from which the motions of the capsid proteins are then determined.

18